\documentstyle[12pt]{article}
\newcommand{\be}{\begin{equation}}
\newcommand{\ee}{\end{equation}}
\newcommand{\bdis}{\begin{displaymath}}
\newcommand{\edis}{\end{displaymath}}

\begin{document}
\centerline{\LARGE Stochastic aggregation model for the multifractal
distribution of matter}
\centerline{\Large F. Sylos Labini$^{1,2}$ and L. Pietronero$^{1}$}
\centerline{\footnotesize ($^1$) Dipartimento di Fisica, Universit\'a di Roma
``La Sapienza'', P.le A.Moro 2,Rome, Italy}
\centerline{\footnotesize ($^2$) Dipartimento di Fisica, Universit\'{a} di Bologna,Italy}
%\maketitle

\section{Introduction}

Two  main features of the observable distribution
of visible matter are the space correlations of galaxy
positions and the mass function of galaxies. As discussed
in Pietronero and Sylos
Labini on this issue ([1], see also [2],[3]),
the concept of multifractal (MF) distribution (including the masses)
naturally unifies these two properties.
Hence with the knowledge of the
whole MF spectrum one obtains information on the
correlations in space as well as on the mass function.

{}From a theoretical point of view one would like to identify
the dynamical process that leads to such a MF distribution.
In order to gain some insight into this complex problem
we have developed
a simple stochastic model that includes some of the basic properties of
aggregation process and allows us to pose a variety of
interesting question concerning the possible dynamical origin
of the MF distribution.
The dynamics is characterised by some parameters,
that have a direct physical meaning in term of cosmological processes.
In this way we can relate the input parameters of the dynamics
to the properties of the final configuration and produce a sort of
phase diagram.

\section{The dynamical friction}
We suppose that the structures are formed
by the aggregation of smaller objects: the main
point which we consider  is that this
process
likely to be dependent on the environment in which it takes place.
If two particles collide, in
order to form a bound state they have to dissipate a certain
amount of energy. The basic mechanism of energy dissipation
via gravitational
interaction is the dynamical friction,
that is a systematic deceleration effect to which a test particle
undergoes moving through a cloud of other particles.
It is a consequence of the fluctuating force acting on the test
particle due to the varying complexion of particles
neighbours, so it is environment dependent and
it is more efficient in more dense regions.
\begin{figure}
\vspace{5cm}
\caption{The probability of making an irreversible aggregation
$\:P_{a}$ during a collision is greater in denser regions (a)
than in sparse ones (b)}
\end{figure}
The dynamical friction, in addition to the local density,
depends also from other quantities:
the relative velocity
and the relative mass of the
test particle with respect to the
velocity and mass of the
background particles [4].
A complete study of this effects as well as the
computation of the typical time scales
of the dynamical friction
in cosmological environment
is still in progress.
In our simulation
when two particles collide there is a probability $\:P_{a}$
of irreversible aggregation and probability $\:1-P_{a}$
to scatter. In such a manner the gravitational interaction is simulated
only through the aggregation probability $\:P_{a}$ that
is made dependent from some parameters,
$\:P_{a} = P_{a}(\alpha,\beta,..)$,
that define the dynamics of the aggregation process.
The environment dependence of the dynamical friction
breaks the spatial
symmetry of the aggregation
process: this is one of the fundamental element
that originates a fractal (and multifractal) distribution.

\section{The simulations}
We consider an aggregation process that growths via binary
collisions between particles. The collision rate is uniform in space
while the probability of forming an irreversible aggregate,
can be a complex function of the local density.
We consider a two dimensional
grid with $\:N=256^{2}$
sites and periodic boundary condition.
At each time step
a new particle of unitary mass is added to the
system and has probability $\:P(k;i,j)$ of
going in the site $\:(i,j)$: Clearly in this model the total
mass is not conserved.
The total number of particles is $\:N = 10^{4}$.
The probability is normalized by the condition:
\be
\sum_{i,j} P(k;i,j) =1
\ee
To study the effect of the non linear dynamics we start
from the trivial case in which the probability constant and equal
in each site. In this case we obtain
the binomial distribution as  mass function: of course
the spatial symmetry of the distribution is not broken
and the homogeneity in space is conserved at all times.

The second step is to consider the aggregation
probability dependent to the
mass of then interacting particles
$\:P(k;i,j) =\frac{1}{A} (1+m(k,i)^{\alpha})$
where $\:A$ is the normalization
constant obtained from eq.(1).
The parameter $\:\alpha$ triggers the effect of the perturbation
on the probability due to the mass present in the
site. If $\:\alpha< 1$ the perturbation is small
and the mass function continues to be bell-shaped.
Otherwise if $\:\alpha > 1$ the mass function
becomes a power law, with the
the exponent that depends from the parameter $\:\alpha$
followed by a the cut-off $\:M^{*}$ that depends from the time $\:k$,
as the mass is not conserved.
Even in this case there is not any
spatial dependence of the merging process so that
the distribution remains homogeneous at all times.

We now consider the environment effect on
the merging phenomenon. A way to estimate the local density,
and mimicking the
dissipation effect
in the simulation,
is to assign an influence function to each particle:
the influence
function of
the
particle in the site $\:(i',j')$ with mass $\:m(k;i',j')$
at the time $\:k$ on the
site $\:(i,j)$ is described by:
\be
f(k;i',j',i,j) = exp(-\frac{d(i',j',i,j)}{m(k;i',j')^{\beta}})
\ee
$\:d(i',j',i,j)$ is the distance between the site $\:(i,j)$
where the collision occurs and the generic site $\:(i',j')$.
The multiparticle influence function is
$\:F(k;i,j) = \sum_{i',j'} f(k;i',j',i,j)$.
We keep the aggregation probability to be
$\:P(k;i,j) = \frac{1}{A} \left( 1 + F(k;i,j)^{\alpha} \right)$
We have introduced two free parameters $\:\alpha$ and $\:\beta$.
For a larger value of  $\:\alpha$ the aggregation occurs in
even denser regions, so that the clustering is stronger and
the fractal dimension lower: this parameter
describes the effect of many particles
influence.
The parameter $\:\beta$
tunes the influence of the size of the mass of each particle:
if $\:\beta$ is enhanced the larger aggregate dominates the smaller.

To study the spatial properties
of the simulation we compute the integral density-density correlation is:
\be
G(\vec{r}) = \int_{0}^{R} d\vec{r} <\rho(\vec{r_{0}})\rho(\vec{r}+\vec{r_{0}})>
 \sim R^{D}
\ee
if $\:D=d=2$ the system is homogeneous otherwise if $\:D<d$ is fractal.

At early times the probability is constant
for each site and the effect of the perturbation due to
the mass distribution is small. For this reason the first
sites are occupied in a random manner. Once a certain mass distribution
has been developed, the energy dissipation mechanism becomes dominant
for aggregation. We can see in the simulation a transient from
an homogeneous distribution towards a fractal behaviour:
there is a rapid increment of the aggregation probability.
\begin{figure}
\vspace{5cm}
\caption{The asymptotic (fractal) distribution for $\:\alpha=3$
and $\:\beta=0.5$}
\end{figure}
\begin{figure}
\vspace{6cm}
\caption{{\em a:}The integrated density-density correlation function:
the final states has dimension $\:D=1.6$ ($\:\alpha=3$
and $\:\beta=0.5$).
{\em b:}: the mass function in the same case}
\end{figure}
The fractal dimension of the final state does not depend from
the initial condition but only from the
parameters of the dynamics: $\:D=D(\alpha,\beta)$.
The mass function of the final distribution is a power law
with an exponential tail.
In a certain region of the phase-space of the dynamics
there is the spatial symmetry breaking of
the aggregation process and
the non linear dynamics generates spontaneously the self-similar
fluctuations of the asymptotic state
thai si an attractive fixed point:
there is not
any dependence from initial conditions:
In the asymptotic time limit (infinite mass)
all the sites will be occupied and the
growth of the aggregates follows
in a self-similar way leading to a MF
distribution: the mass function is
Press-Schecheter like.
The breaking of the spatial symmetry in the aggregation probability
will lead  asymptotically to region that will never filled,
the voids, because the density inside is
lower, and with the time evolution
always depressed, than the density in cluster.

\section{Conclusion}
We consider an aggregation process
in which the formation of structure is a process that depends from
{\em the local environment}. The energy dissipation
mechanism through the gravitational
interaction is the dynamical friction, that strongly depends from the
local density. The aggregation is an environment dependent
 process and this breaks spontaneously
the spatial symmetry with respect to the formation of
strictures.
We have shown that in this model the asymptotic distribution
is not generated by an amplification of the initial
density fluctuations but
the non linear dynamics leads spontaneously the
self-similar fluctuations of the asymptotic state, so
that there is not any crucial dependence from initial conditions.
The necessary ingredients for a dynamics in order to generate a fractal
(multifractal) distribution are the breaking of the spatial symmetry,
and the Self-Organized nature of the dynamical mechanism.

\section*{References}
\begin{itemize}
\item [\mbox{[1]}] Pietronero, L., \& Sylos Labini, F. preprint
\item [\mbox{[2]}] Coleman, P.H. \& Pietronero, L.,1992 Phys.Rep. 231,311
\item [\mbox{[3]}] Pietronero, L., \& Tosatti, E. (eds) 1986, Fractals in
Physics
\item [\mbox{[4]}] Chandrasekhar, S. 1943 Principles of stellar dynamics ,Dover
(New York)
\end{itemize}

\end{document}